\documentclass[journal]{IEEEtran}

\usepackage{graphicx}
\usepackage{color}
\usepackage{mathrsfs}
\usepackage{amsmath}
\usepackage{multirow}
\usepackage{amsfonts}
\usepackage[switch]{lineno}

\begin{document}

\title{A Model Compression Method with Matrix Product Operators for Speech Enhancement}

\author{Xingwei~Sun$^{\star}$, Ze-Feng~Gao$^{\star}$, Zhong-Yi~Lu, Junfeng~Li and Yonghong~Yan
\thanks{Xingwei Sun is with Key Laboratory of Speech Acoustics and Content Understanding, Institute of Acoustics, Chinese Academy of Sciences and University of Chinese Academy of Sciences, Beijing, 100190, China.}
\thanks{Ze-Feng Gao is with School of Science, Renmin University of China, Beijing, 100872, China.}
\thanks{Zhong-Yi Lu is with School of Science, Renmin University of China, Beijing, 100872, China (e-mail: zlu@ruc.edu.cn).}
\thanks{Junfeng Li is with Key Laboratory of Speech Acoustics and Content Understanding, Institute of Acoustics, Chinese Academy of Sciences and University of Chinese Academy of Sciences, Beijing, 100190, China (e-mail: lijunfeng@hccl.ioa.ac.cn).}
\thanks{Yonghong Yan is with Key Laboratory of Speech Acoustics and Content Understanding, Institute of Acoustics, Chinese Academy of Sciences and University of Chinese Academy of Sciences, Beijing, 100190, China. He is also with Xinjiang Key Laboratory of Minority Speech and Language Information Processing, Xinjiang Technical Institute of Physics and Chemistry, Chinese Academy of Sciences.}
\thanks{$^{\star}$ Refers to an equal contribution.}}

\maketitle

\begin{abstract}
The deep neural network (DNN) based speech enhancement approaches have achieved promising performance. However, the number of parameters involved in these methods is usually enormous for the real applications of speech enhancement on the device with the limited resources. This seriously restricts the applications. To deal with this issue, model compression techniques are being widely studied. In this paper, we propose a model compression method based on matrix product operators (MPO) to substantially reduce the number of parameters in DNN models for speech enhancement. In this method, the weight matrices in the linear transformations of neural network model are replaced by the MPO decomposition format before training. In experiment, this process is applied to the causal neural network models, such as the feedforward multilayer perceptron (MLP) and long short-term memory (LSTM) models. Both MLP and LSTM models with/without compression are then utilized to estimate the ideal ratio mask for monaural speech enhancement. The experimental results show that our proposed MPO-based method outperforms the widely-used pruning method for speech enhancement under various compression rates, and further improvement can be achieved with respect to low compression rates. Our proposal provides an effective model compression method for speech enhancement, especially in cloud-free application.
\end{abstract}


%
\IEEEpeerreviewmaketitle

\section{Introduction}

%
%
%
\IEEEPARstart{S}{peech} enhancement techniques to mitigate the harmful effects of background noise and interference have been studied for several decades with a variety of promising applications, such as telecommunications and hearing aid systems~\cite{loizou2007speech}. Monaural speech enhancement is widely used to improve the speech quality and speech intelligibility. Conventional speech enhancement approaches exploit the different characteristics of speech and noise in the time-frequency (T-F) domain to suppress the background noise, such as spectral subtraction~\cite{boll1979suppression}, Wiener filter~\cite{Lim1979Enhancement}, and statistical model-based estimators~\cite{ephraim1984speech}. However, the performance of these methods often greatly degrades especially in non-stationary noise conditions.

In recent years, the deep neural network (DNN)-based speech enhancement methods have achieved promising performance, especially in non-stationary noise conditions~\cite{wang2018supervised}. The DNN-based monaural speech enhancement methods are mainly divided into the mapping-based and masking-based approaches. The mapping-based approaches adopt DNN as a regression model to directly map the log power spectrum of noisy speech to that of clean speech~\cite{xu2014experimental}~\cite{xu2015regression}. In the masking-based methods, the mask in the T-F domain is estimated and further applied to the noisy spectrum. In~\cite{wang2013towards}, DNNs are first trained as binary classifiers to predict the ideal binary mask (IBM) for removing the background noise. The IBM is proposed based on the concept of computational auditory scene analysis (CASA)~\cite{wang2005ideal}, which assigns 1 to a T-F unit if the speech energy within the unit exceeds the noise energy and 0 otherwise. Subsequently, the ideal ratio mask (IRM)~\cite{wang2014training} which assigns a soft label between 0 and 1 rather than binary label to each T-F unit and yields better speech quality. More recently, the masks involving phase information have been proposed to improve the enhancement performance, such as the phase-sensitive mask (PSM)~\cite{erdogan2015phase} and complex ideal ratio mask (cIRM)~\cite{williamson2016complex}.

Deep learning techniques, especially deep neural networks, have developed rapidly in recent years. In DNN-based speech enhancement approaches, the feedforward multilayer perceptrons (MLPs) are first used for spectral mapping and mask estimation. For MLPs, a window of consecutive time frames is typically utilized to provide the temporal contexts information. Without the ability of leveraging long term information, the generalization ability of the MLP-based approaches to speakers is limited~\cite{kolbk2017speech}. Speech enhancement is formulated as a sequence-to-sequence mapping in~\cite{chen2017long} and recurrent neural network (RNN) with long short-term memory (LSTM) layers is utilized to address speaker generalization. RNNs have been proven to perform better than MLPs for speech enhancement~\cite{weninger2014single}~\cite{weninger2015speech}. Convolutional neural network (CNN) has also been used for speech enhancement recently. In~\cite{park2016fully}, a convolutional encoder-decoder network (CED) is used to learn a spectral mapping function and then shows the similar denoising performance compared with the MLP and RNN model with much smaller model size. Afterwards,  a fully convolutional network (FCN) is proposed for straightforward mapping from a noisy waveform to the corresponding clean waveform~\cite{fu2017raw}. In~\cite{pandey2019tcnn}, an additional temporal convolutional module using causal and dilated convolutional layers is inserted between the encoder and the decoder of the CED-based speech enhancement architecture. This method obtained better performance with much fewer parameters benefited from the dilated convolution layers compared with an LSTM-based method.

Thus far, the DNN-based speech enhancement approaches have achieved good performance by designing complicated neural network architectures. However, these methods often involve enormous number of parameters with high memory requirements.
To address the compromise between high-performance and compact model size, lots of notable works have been done.
Knowledge distilling is proposed and successfully applied in MLPs~\cite{hinton2015distilling} and CNNs~\cite{ba2014deep}, which compresses a large deep neural network into a smaller neural network by training the latter model on the transformed softmax outputs from the former. This also works for RNNs~\cite{tang2016recurrent}.
With more simplicity, the pruning algorithm is also proposed to reduce the model size~\cite{han2015deep}~\cite{han2015learning}, in which the network is pruned to learn the main important connections through removing the connections with small weights. The pruned model can be retrained to fine tune the remaining connections. The weight matrices of the pruned model can be considered sparse which means the model can be compressed by only keeping the non-zero values and their corresponding coordinates.
Weight sharing and quantization methods assume that many weights of a well trained model have similar values. Thus, the weight values can be grouped with grouping methods such as hashing~\cite{chen2015compressing}, k-means, and vector quantization~\cite{han2015deep}, in order to reduce the number of free parameters.
In~\cite{xue2013restructuring}, a singular value decomposition (SVD) approach was applied to the weight matrices of a well-trained MLP model. Then the model is restructured based on the inherent sparseness of the original matrices which reduce the model size significantly with negligible accuracy loss while the model can also be retrained to lessen the accuracy loss. Low-rank decomposition is also used for model size compression by replacing the weight matrices in fully-connected layers with their low-rank approximations, obtained with truncated SVD~\cite{denil2013predicting}. Similar to the low-rank approximation, the Tensor-Train format~\cite{oseledets2011tensor} is used to replace the dense weight matrices of fully-connected layers~\cite{novikov2015tensorizing} and embedding layers~\cite{Khrulkov2015tensorized}. In~\cite{gao2019compressing}, the matrix product operators (MPO) decomposition format developed from quantum many-body physics and based on high order tensor single value decomposition method, is proved to be well effective for model compression by replacing the linear transformations of fully-connected and convolutional layers. Instead of replaced the CNN layers with the MPO format in~\cite{gao2019compressing}, the weight matrix of each convolution kernel in CNN layer is decomposed with MPO format to compress the model with CNN layers in~\cite{garipov2016ultimate}. In~\cite{Yinchong2017tensor}, the compression method for recurrent neural network was proposed in a video classification task. However, only the input-to-hidden weight matrix in LSTM model has been compressed with the Tensor-Train format and the hidden-to-hidden weight matrix remain unchanged. After these progresses for model size compression, quantization and Huffman coding can be used to further reduce the memory requirement~\cite{han2015deep}.

In real applications of speech enhancement, such as speech communication, the memory resource is restricted. Therefore, in addition to achieve good performance, the compact model size is also important for DNN-based speech enhancement approaches. In this paper, we focus on reducing the model size before using quantization and Huffman coding. We propose a model compression method based on the MPO decomposition and apply it to the casual neural network models, such as MLP and LSTM, for speech enhancement. In this method, the linear transformations in MLP and LSTM models are replaced by MPO decomposition format. In experiment, both MLP and LSTM models with and without compression are trained to estimate the IRM in various noise environments. The speech enhancement performance of the compressed model using our proposed MPO-based method is evaluated and compared with the pruning method under the compression rates from 5 to 100 times. The experimental results show that, the MPO-based method outperforms the pruning method for two DNN models under the same compression rates and satisfactory performance can still be obtained with small model size with MPO-based model compression. Moreover, further improvement has been achieved in low compression rates with the MPO-based method.
Overall, our proposed MPO-based model compression method for speech enhancement has threefold benefits: (1) we applied the MPO-based decomposition for both the input-to-hidden and hidden-to-hidden weight matrixes in LSTM model which is different from the previous MPO-based compression methods; (2) this method outperforms the pruning method in speech enhancement performance while the satisfactory performance can still be obtained with small model size, and further improvement can be obtained at low compression rate; (3) this MPO-based method can be easily integrated to any network models with linear transformations and trained with the other parts of the model.

The rest of this paper is organized as follows. The DNN-based monaural speech enhancement system is introduced in Section \ref{sec:problem}. In Section \ref{sec:mpo}, we describe the MPO-based model compression method in detail. Experimental setup is provided in Section \ref{sec:experiment}, followed by the experimental results and analysis in Section \ref{sec:result}. Finally, the discussion and conclusion are presented in Section \ref{sec:conclusion}.

\section{DNN-based Monaural Speech Enhancement}
\label{sec:problem}

\begin{figure}
\centering
\includegraphics[width= \linewidth]{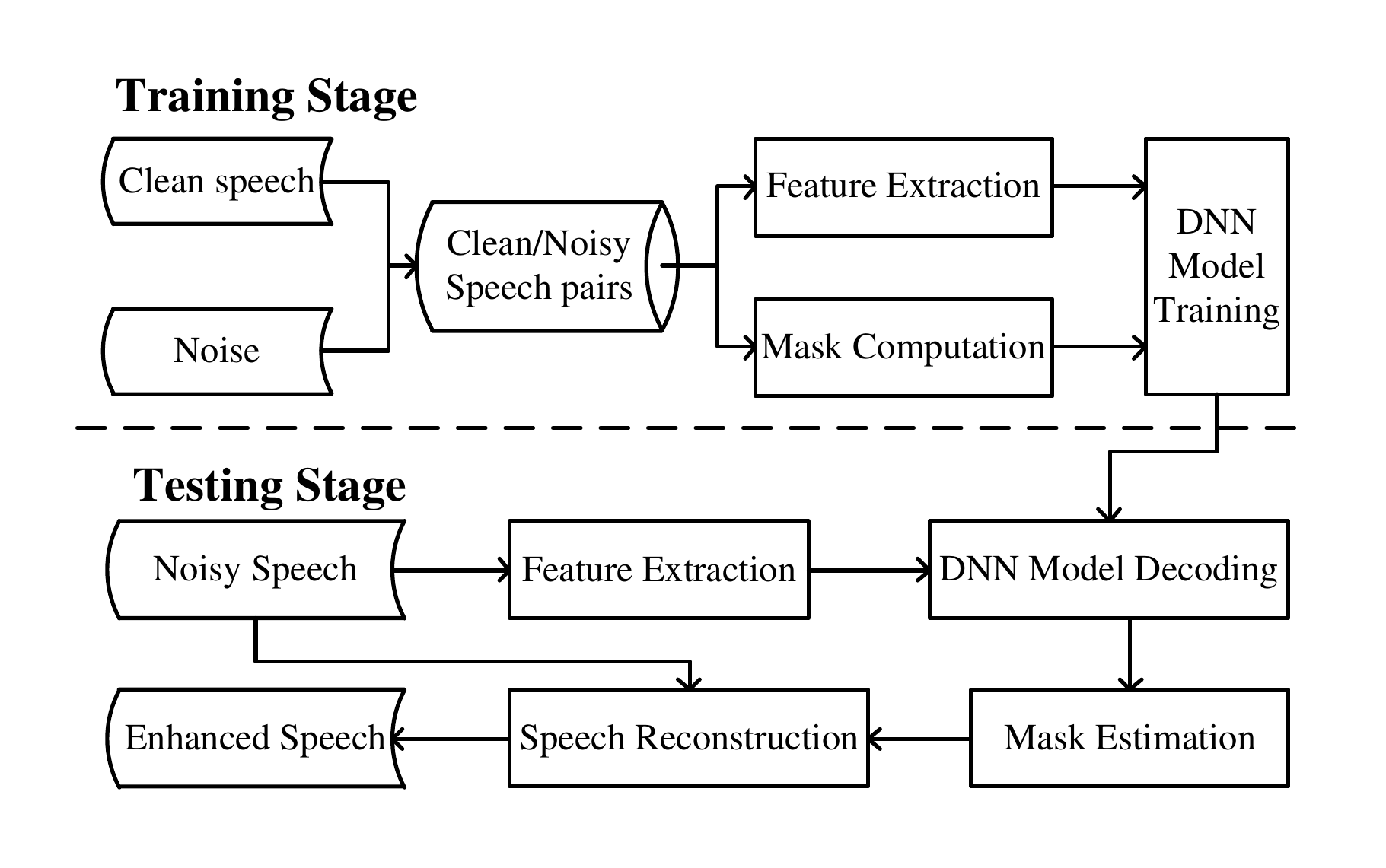}
\caption{A block diagram of monaural speech enhancement framework with DNN-based mask estimation.}
 \label{figse}
\end{figure}
 The block diagram of the monaural speech enhancement framework with DNN-based mask estimation is illustrated in Fig.\ref{figse}. In the training stage, the DNN-based model is trained using the acoustic features and masking-based target from pairs of clean and noisy speech data. In the testing stage, the well-trained DNN model is fed with the acoustic features of noisy speech for mask estimation, while  the enhanced speech is reconstructed with the noisy speech and estimated mask. More details about the framework used in this paper are presented in the following subsections.

\subsection{Signal Model}

In noisy environments, speech signal is often corrupted by background noise. The received noisy speech signal $y(k)$ at a microphone can be described as:
\begin{equation}\label{eq:signal_model_time}
      y(k) =s(k)+n(k),
\end{equation}
where $k$ indicates the time sample index, $s(k)$ and $n(k)$ denote the target speech and additive background noise signals, respectively. The objective of speech enhancement is to remove the noise signal from the received signal and retain the target speech signal as much as possible.
As the speech enhancement methods usually process the signals in frequency domain, the signal model in Eq.(\ref{eq:signal_model_time}) can be transformed into frequency domain with short-time Fourier transform (STFT) and described as:
\begin{equation}\label{eq:signal_model_freq}
     Y(l,f) =S(l,f)+N(l,f),
\end{equation}
where $l$ and $f$ indicate the time frame index and frequency bin index, and $Y(l,f)$, $S(l,f)$ and $N(l,f)$ denote the complex spectrum of the corresponding signals after STFT process.

\subsection{Acoustic Features and Training Target}

In the training stage, the acoustic features and the masking-based target are computed from the clean and noisy speech pairs. Then the acoustic features and masking-based target are fed into the DNN model for training. In the testing stage, the acoustic features extracted from the noisy speech are fed into the well-trained model for mask estimation.

The effect of different acoustic features of DNN-based speech enhancement have been studied in many researches. Inspired by~\cite{wang2013exploring}, we combine several acoustic features to form a complementary feature set for the DNN model training. This complementary feature set includes the amplitude modulation spectrogram (AMS), relative spectral transform and perceptual linear prediction (RASTA-PLP), mel-frequency cepstral coefficients (MFCC), cochleagram response and their deltas. The acoustic features of each frame are computed.

To demonstrate the effectiveness of the proposed model, we use the models to estimate the IRM, which is a widely-used masking-based training target in DNN-based speech enhancement. It can be computed as:
\begin{equation}\label{eq:irm}
    IRM(l,f)=\sqrt{ \frac{|S(l,f)|^2}{|S(l,f)|^2 + |N(l,f)|^2}}.
\end{equation}
The value of IRM is between 0 and 1. In the testing stage, the estimated IRM is multiplied by the magnitude of the noisy speech signal and the noisy phase is used for the signal reconstruction with inverse STFT (ISTFT) process.

\subsection{DNN Model Architecture}
\label{subsec:model}

In this section, the basic architectures of MLP and LSTM are described. Both models are used for mask estimation in this study.

\subsubsection{Basic Architecture of  MLP Model}

The MLP model is comprised of multiple fully-connected layers. Without loss of generality, we use the forward progress of the first layer as an example which can be described as:
\begin{equation}\label{eq:mlp}
    \textbf{h} = F(\textbf{W}\textbf{x} + \textbf{b}),
\end{equation}
where $\textbf{h}$ and $\textbf{x}$ are the value of hidden and input layer units, $\textbf{W}$ and $\textbf{b}$ are the trainable weight matrix and bias, and $F$ is the activation function. The MLP model can be trained with the back propagation algorithm.

\subsubsection{Basic architecture of LSTM Model}

As the speech signal has strong temporal correlation, RNN model with LSTM layers is used for learning the temporal dynamics of speech.
An LSTM block has a memory cell and three gates in which the input gate controls how much information should be added to the cell, the forget gate controls how much previous information should be erased from the cell, and the output gate controls how much information should be transformed to the next layer. In this study, the LSTM is defined by the following equations:
\begin{eqnarray}
        \label{eq:lstm1}
        &\textbf{i}^{[t]}&= \sigma(\textbf{W}^i\textbf{x}^{[t]}+\textbf{U}^i\textbf{h}^{[t-1]}+\textbf{b}^i),  \\
        \label{eq:lstm2}
        &\textbf{f}^{[t]}&= \sigma(\textbf{W}^{f}\textbf{x}^{[t]}+\textbf{U}^{f}\textbf{h}^{[t-1]}+\textbf{b}^{f}),  \\
        \label{eq:lstm3}
        &\textbf{o}^{[t]}&= \sigma(\textbf{W}^{o}\textbf{x}^{[t]}+\textbf{U}^{o}\textbf{h}^{[t-1]}+\textbf{b}^{o}),  \\
        \label{eq:lstm4}
        &\textbf{g}^{[t]}&= tanh(\textbf{W}^{g}\textbf{x}^{[t]}+\textbf{U}^{g}\textbf{h}^{[t-1]}+\textbf{b}^{g}),  \\
        &\textbf{c}^{[t]}&= \textbf{f}^{[t]} \otimes \textbf{c}^{[t-1]} + \textbf{i}^{[t]} \otimes \textbf{g}^{[t]},  \\
        &\textbf{h}^{[t]}&= \textbf{o}^{[t]} \otimes tanh(\textbf{c}^{[t]}),
 \end{eqnarray}
where $\textbf{x}^{[t]}$, $\textbf{g}^{[t]}$, $\textbf{c}^{[t]}$, $\textbf{h}^{[t]}$ represent input, block input, memory cell, and hidden activation output at current time step $t$, respectively. $\textbf{c}^{[t-1]}$ and $\textbf{h}^{[t-1]}$ represent memory cell and hidden activation output at previous time step $t-1$. The input gate, forget gate, and output gate are denoted as $\textbf{i}$, $\textbf{f}$, and $\textbf{o}$ respectively. $\sigma$ and $tanh$ represent the sigmoid and hyperbolic tangent activation functions respectively. $\otimes$ denotes element-wise multiplication. $\textbf{W}$, $\textbf{U}$, and $\textbf{b}$ with different superscripts are the input-to-hidden, hidden-to-hidden weight matrices, and bias in different gates and the block input respectively.

\section{MPO-based Model Compression}
\label{sec:mpo}

In this section, we first give an introduction to the core ingredient of the MPO decomposition format which is the basis of MPO-based model compression method. Then we replace the linear transformations in MLP and LSTM models by an MPO decomposition format to obtain compact models for monaural speech enhancement. Note that, the MPO representation can be trained end-to-end, thus it can work together with the rest of models in a very efficient way and no extra process is needed for model training.

\subsection{Matrix Product Operators}
\label{subsec:mpos}

\begin{figure*}
    \centering
    \includegraphics[width=0.9\linewidth]{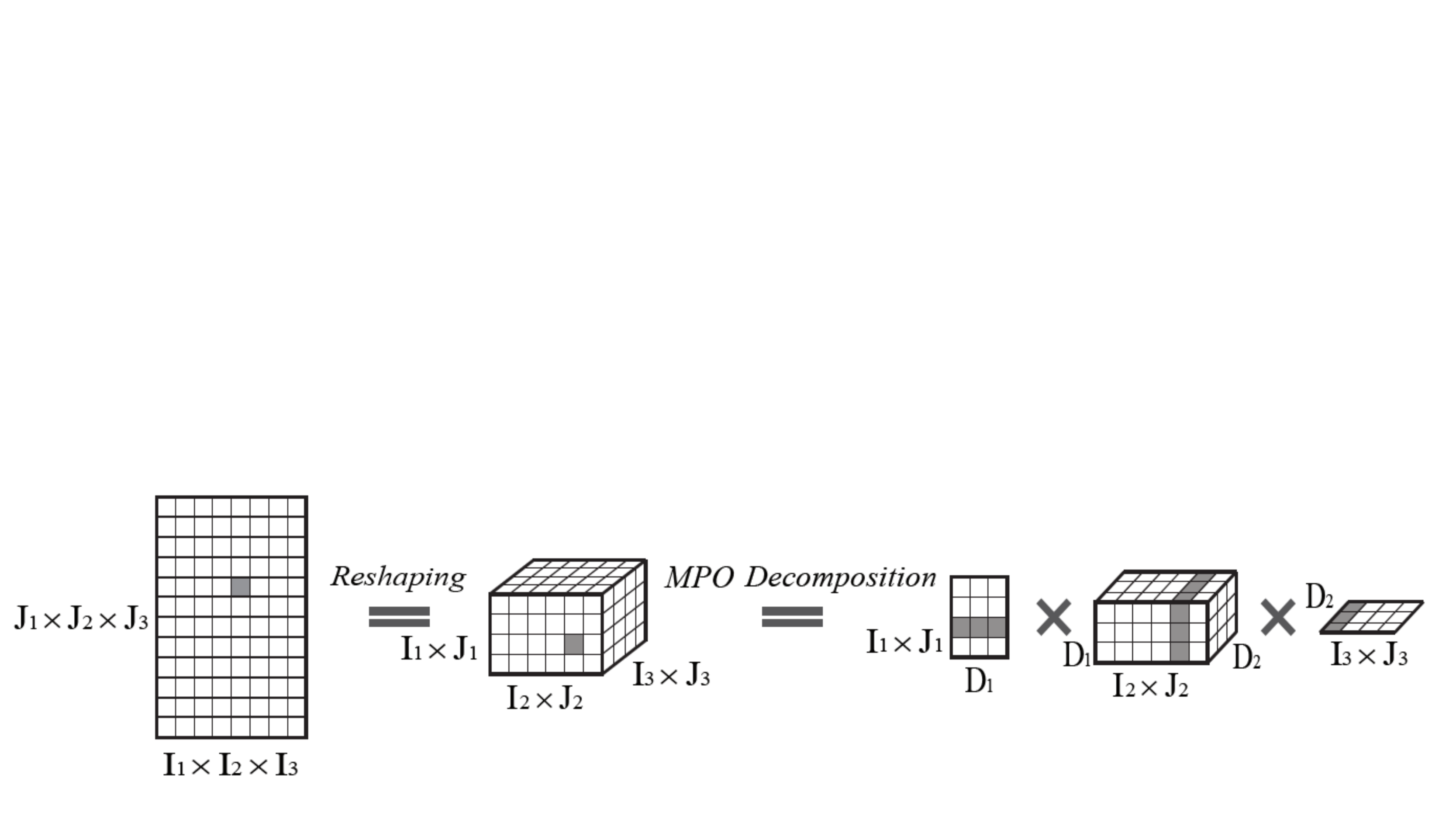}
    \caption{Construction of MPO decomposition format matrices from a standard weight matrix. Gray color depicts how a single element in an initial matrix is transformed into a high-dimension matrix and represented with the product of a sequence of matrices in MPO decomposition format.}
    \label{fig_mpo}
\end{figure*}

An MPO decomposition format, developed from quantum many-body physics,  is a more generalized form of the Tensor-Train format that is used to factorize a higher-order tensor into a sequential product of the so-called local-tensors~\cite{verstraete2004matrix}. By representing the linear transformations in a model with MPO decomposition format, the number of parameters requirement is greatly shrunk since the number of parameters contained in an MPO decomposition format just grows linearly with the system size\cite{poulin2011quantum}.

To clarify the MPO decomposition process, we assume that a weight matrix $\textbf{W}_{I,J} \in\mathbb{R}^{I\times J}$ is matrix with size $I\times J$. Given two arbitrary factorizations of its dimensions into natural numbers, we can reshape and transpose this matrix into an $N$-dimension tensor $\textbf{W}_{I_1\cdots I_k\cdots I_N,J_1\cdots J_k\cdots J_N}\in\mathbb{R}^{I_1J_1\times\cdots \times I_kJ_k\times\cdots\times I_NJ_N}$, in which:
\begin{equation} \label{eq:dims}
    \prod_{k=1}^{N}I_{k} = I, \quad \prod_{k=1}^{N}J_{k} = J.
\end{equation}
This process can be regarded as one-to-one element mapping of a low-dimension matrix to a high-dimension tensor. Then the MPO decomposition can be applied to a high-dimension tensor and leading to a more compact representation. A high-dimension tensor can be represented as the product of a sequence of matrices which are called local-tensors.
A visualized description of the MPO decomposition for a 3-dimension tensor which is mapped from a standard matrix is shown in Fig.\ref{fig_mpo}~\cite{Khrulkov2015tensorized}. The gray color depicts that how a single element in an initial matrix is transformed into a high-dimension matrix and represented with the product of local-tensors. This decomposition can be written as:
\begin{equation}  \label{eq:mpodc}
    \begin{aligned}
          \textbf{W}_{I_1I_2\cdots I_N,J_1J_2\cdots J_N} &= \\
               \textbf{W}^{(1)} [I_1,J_1] & \cdots \textbf{W}^{(k)} [I_k,J_k] \cdots \textbf{W}^{(N)} [I_N,J_N],
    \end{aligned}
\end{equation}
where $\textbf{W}^{(k)} [I_k,J_k]$ is a 4-dimensional tensor with size $ D_{k-1}\times I_k\times J_k \times D_{k}$ in which $D_k$ is a bond dimension linking $\textbf{W}^{(k)}$ and $\textbf{W}^{(k+1)}$ with $D_0=D_N=1$.
The factorization factors, $I_k$ and $J_k$, and the bond dimension, $D_{k}$, of a local tensor are the tunable hyper-parameters of a compact model to get the given compression rate. Empirically, the number of local tensor would be increased and the bond dimension would be reduced to obtain lager compression rate in a network model.

Such a matrix product structure in an MPO representation leads to such a fact that the scaling of the parameter number with dimensions of a tensor is reduced from exponential to polynomial, which is a great advantage of MPO representation. To be specific, the parameter number is shrunk with the following equation:
\begin{equation}\label{eq:reduced_D}
   \prod_{k=1}^{N}I_k J_k \rightarrow \sum_{k=1}^{N}I_k J_k D_{k-1}D_k.
\end{equation}
When $\textbf{W}^{(k)}[I_k,J_k]$ is of full rank, which means $D_k = I_k J_k D_{k-1}$, there is no parameter shrinking. Therefore, by reducing the dimension of $D_k$, we can get rid of the redundant parameters to compress the model. In practice, ${D_k}$ is tuned to get a tradeoff between model compression and fitting ability.

\subsection{MPO-based Compression of MLP Model}
\label{sec:mpomlp}

As introduced in subsection \ref{subsec:model}, the core ingredient of MLP model is its fully-connected layer. It can be regarded as a linear transformation without considering the activation function. Assuming the input and hidden layers in MLP model have $N_x$ and $N_h$ units, Eq.(\ref{eq:mlp}) can be written as:
\begin{equation}\label{eq:linear}
    \textbf{h}_{N_h} = F(\textbf{W}_{N_h,N_x}\textbf{x}_{N_x} + \textbf{b}_{N_h}).
\end{equation}
In the linear transformation part, the parameters in weight matrix $\textbf{W}_{N_h,N_x}\in\mathbb{R}^{N_h\times N_x}$ can be compressed using MPO decomposition. Specifically, with $\prod_{k=1}^{N}I_{k} = N_h, \prod_{k=1}^{N}J_{k} = N_x$, the weight matrix $\textbf{W}_{N_h,N_x}$ can be written as:
\begin{equation}  \label{eq:mpomlp}
    \begin{aligned}
    \textbf{W}_{N_h,N_x} &=   \textbf{W}_{I_1\cdots I_k \cdots I_N,J_1\cdots J_k\cdots J_N} \\
               &=\textbf{W}^{(1)} [I_1,J_1] \cdots \textbf{W}^{(k)} [I_k,J_k] \cdots \textbf{W}^{(N)} [I_N,J_N],
    \end{aligned}
\end{equation}
where $\textbf{W}^{(k)} [I_k,J_k]$ is a 4-dimensional tensor with size $ D_{k-1}\times I_k\times J_k \times D_{k}$ and these tunable hyper-parameters are adjusted to get a given compression rate.
After the MPO decomposition of the weight matrix, the linear transformation can be written as:
\begin{equation}\label{eq:mpo}
    \begin{aligned}
    &MPO(\textbf{W}_{N_h,N_x}, \textbf{x}_{N_x}) = \\
    &\textbf{W}^{(1)} [I_1,J_1] \cdots \textbf{W}^{(k)} [I_k,J_k] \cdots \textbf{W}^{(N)} [I_N,J_N]\textbf{x}_{J_1\cdots J_k\cdots J_N},
    \end{aligned}
\end{equation}
where $\textbf{x}_{J_1\cdots J_k\cdots J_N}$ is the input tensor reshape and transpose into an N-dimension from $\textbf{x}_{N_x}$.
Finally, we term the forward progress of the fully-connected layer as:
\begin{equation}\label{eq:mpolinear}
    \textbf{h}_{N_h} = F(MPO(\textbf{W}_{N_h,N_x}, \textbf{x}_{N_x}) + \textbf{b}_{N_h}),
\end{equation}
%
Before the model training, the weight matrices are replaced with MPO decomposition format. As the MPO representation can be trained with the other parts of the model, the back propagation algorithm can still be used for training. The parameter compression rate of one fully-connected layer in MLP model is given by:
\begin{equation}\label{eq:mlprho}
    \rho_{MLP} = \frac{\sum_{k=1}^{N}I_{k}J_{k}D_{k-1}D_{k}+N_h}{N_hN_x+N_h}.
\end{equation}

\subsection{MPO-based Compression of LSTM Model}
\label{subsec:mpllstm}

As linear transformations are the basic operators of a LSTM model, it can also be compressed using MPO-based method.
Inspired by~\cite{chollet2015keras}, we reorganize the basic operators of LSTM model to obtain better compression performance. Take the first layer of LSTM model as an example with $N_x$ input and $N_h$ hidden units, the basic operators presented in Eqs.(\ref{eq:lstm1})-(\ref{eq:lstm4}) can be written as:

\begin{equation}\label{eq:relstm}
    \textbf{m}^{[t]}_{4N_h} = F(  (\textbf{W}_{4N_h,N_x} \textbf{x}^{[t]}_{N_x} + \textbf{U}_{4N_h,N_h} \textbf{h}^{[t-1]}_{N_h}  ) +\textbf{b}_{4N_h} ),
\end{equation}
where
\begin{equation}\label{eq:mfb}
    \textbf{m}_{4N_h} = \begin{pmatrix} \textbf{i}^{[t]}_{N_h}\\ \textbf{f}^{[t]}_{N_h}\\ \textbf{o}^{[t]}_{N_h}\\ \textbf{g}^{[t]}_{N_h}\\ \end{pmatrix},
    F=\begin{pmatrix} \sigma \\ \sigma\\ \sigma\\ \tanh \\ \end{pmatrix},
   \textbf{b}_{4N_h}=\begin{pmatrix} \textbf{b}^i_{N_h}\\ \textbf{b}^f_{N_h}\\ \textbf{b}^o_{N_h}\\ \textbf{b}^g_{N_h} \end{pmatrix},
\end{equation}
\begin{equation}\label{eq:WU}
    \textbf{W}_{4N_h,N_x}=\begin{pmatrix}\textbf{W}^i_{N_h,N_x}\\ \textbf{W}^f_{N_h,N_x}\\ \textbf{W}^o_{N_h,N_x}\\ \textbf{W}^g_{N_h,N_x} \end{pmatrix},
    \textbf{U}_{4N_h,N_h}=\begin{pmatrix} \textbf{U}^i_{N_h,N_h}\\ \textbf{U}^f_{N_h,N_h}\\ \textbf{U}^o_{N_h,N_h}\\ \textbf{U}^g_{N_h,N_h} \end{pmatrix}.
\end{equation}
Then in the linear transformation parts of Eq.(\ref{eq:relstm}), the parameters in input-to-hidden weight matrix $\textbf{W}_{4N_h,N_x}\in \mathbb{R}^{4N_h\times N_x}$ and hidden-to-hidden weight matrix $\textbf{U}_{4N_h,N_h}\in \mathbb{R}^{4N_h\times N_h}$ can be compressed using MPO decomposition format. With $\prod_{k=1}^{N_W}I_{k}^W = 4N_h, \prod_{k=1}^{N_W}J_{k}^W = N_x, \prod_{k=1}^{N_U}I_{k}^U = 4N_h$ and $\prod_{k=1}^{N_U}J_{k}^U = N_h$, we can have:
\begin{equation}  \label{eq:mpow}
    \begin{aligned}
        &\textbf{W}_{4N_h,N_x}=   \textbf{W}_{I_1^W\cdots I_k^W \cdots I_{N_W}^W,J_1^W\cdots J_k^W\cdots J_{N_W}^W} \\
        &=\textbf{W}^{(1)} [I_1^W,J_1^W] \cdots \textbf{W}^{(k)} [I_k^W,J_k^W] \cdots \textbf{W}^{({N_W})} [I_{N_W}^W,J_{N_W}^W],
    \end{aligned}
\end{equation}
and
\begin{equation}  \label{eq:mpou}
    \begin{aligned}
        &\textbf{U}_{4N_h,N_h}=   \textbf{U}_{I_1^U\cdots I_k^U \cdots I_{N_U}^U,J_1^U\cdots J_k^U\cdots J_{N_U}^U} \\
        &=\textbf{U}^{(1)} [I_1^U,J_1^U] \cdots \textbf{U}^{(k)} [I_k^U,J_k^U] \cdots \textbf{U}^{({N_U})} [I_{N_U}^U,J_{N_U}^U],
    \end{aligned}
\end{equation}
where $\textbf{W}^{(k)} [J_k^W,J_k^W]$ is a 4-dimensional tensor with size $ D_{k-1}^W \times I_k^W\times J_k^W \times D_{k}^W$ and $\textbf{U}^{(k)} [J_k^U,J_k^U]$ with size $ D_{k-1}^U \times I_k^U\times J_k^U \times D_{k}^U$.
After that, Eq.(\ref{eq:relstm}) can be written in MPO format as:
\begin{equation}\label{eq:mpolstm}
    \begin{aligned}
        \textbf{m}^{[t]}_{4N_h}= F (  ( &MPO( \textbf{W}_{4N_h,N_x}, \textbf{x}^{[t]}_{N_x}) \\
        + &MPO(\textbf{U}_{4N_h,N_h}, \textbf{h}^{[t-1]}_{N_h}) ) +\textbf{b}_{4N_h}  ).
    \end{aligned}
\end{equation}

Finally, the compression rate of the input-to-hidden weight matrix $\textbf{W}_{4N_h,N_x}$ is:
\begin{equation}
    \rho_{W} = \frac{\sum_{k=1}^{N_W}I_{k}^WJ_{k}^WD_{k-1}^WD_{k}^W}{4{N_h}N_x},
\end{equation}
and the compression rate of the hidden-to-hidden weight matrix $\textbf{U}_{4N_h,N_h}$ is:
\begin{equation}
    \rho_{U} = \frac{\sum_{k=1}^{N_U}I_{k}^UJ_{k}^UD_{k-1}^UD_{k}^U}{4{N_h}N_h}.
\end{equation}
Thus, the compression rate of one layer in LSTM model can be computed as:
\begin{equation}
    \begin{aligned}
        &\rho_{LSTM} = \\
        &\frac{\sum_{k=1}^{N_W}I_{k}^WJ_{k}^WD_{k-1}^WD_{k}^W + \sum_{k=1}^{N_U}I_{k}^UJ_{k}^UD_{k-1}^UD_{k}^U + 4N_h}{4{N_h}N_x+4{N_h}N_h+ 4N_h}.
    \end{aligned}
\end{equation}

\section{Experimental Setup}
\label{sec:experiment}

\subsection{Data Preparation}
\label{subsec:data}
In the experiments, we used a Chinese speech corpus which consists of speech utterances from a number of male and female speakers. We randomly selected 4620 speech utterances used for training and 100 speech utterances used for testing. The speech utterances used for testing are not included in the training set.
To obtain noise-independent model, 20 types of noises from the NOISEX92~\cite{Andrew1993Assessment}, DEMAND~\cite{thiemann2013DEMAND} and CHiME-III~\cite{Barker2015CHiME} corpora were used and divided into 3 noise datasets respectively, as shown in Table \ref{tab-noise}.
\begin{table}[!htb]
\caption{Noise datasets used in this experiment.}
\label{tab-noise}
\centering \renewcommand{\arraystretch}{1.1}
  \begin{tabular}{c|c}  \hline
        Noise Dataset & Noise Type \\ \hline
        \multirow{4}{*}{A} &  buccaneer2, f16, factory1, volvo,\\  & PCAFETER, PRESTO, PSTATION, TCAR \\    & SPSQUARE, SCAFE,STRAFFIC, TBUS\\    \hline
        B &  babble, pink,  leopard, white \\ \hline
        C &  bus, cafe, pedestrian, street \\    \hline
 \end{tabular}
\end{table}

We evaluated both MLP and LSTM models in two different monaural speech enhancement tasks. In \emph{Task \uppercase\expandafter{\romannumeral1}}, only noise dataset C was used with the first half of each noise signal for training and the second half for testing. In \emph{Task \uppercase\expandafter{\romannumeral2}}, the noise datasets A and B were used for training,while the noise dataset C was used for testing. The same speech datesets were used for both tasks. When generating the noisy speech signals, the clean speech signals and noise signals were mixed at three different signal-to-noise ratio (SNR) levels ($-$5 dB, 0 dB and 5 dB). We expanded a speech utterance shorter than 5 seconds to 5 seconds by repeating it or randomly truncating the one longer than 5 seconds before mixing with the noise that was expanded or truncated in the same way. We generated 55440 and 1200 noisy speech utterances in training and testing datasets respectively for each task by randomly selecting the noise types for each speech utterance, which are about 77 and 1.7 hours, respectively.

\subsection{Baseline and Training Details}

The signal at 16 kHz sample rate was framed using a 32 ms Hamming window with a 16 ms window shift. For each frame, a 512 point STFT was performed and then the IRM training target with 257 dimensions was generated. In order to be consistent with the compression methods, the first dimension of IRM was discarded to obtain a 256 dimension training target and it was padded back with zeros in the testing stage. The dimension of the complementary acoustic features is also 256. The zero mean and unit variance normalization are applied to the input features with the global mean and variance values before feeding into the models in both training and testing stage. Each dimension of the global mean and variance values is computed from the corresponding dimension of the input features in the training dataset.

In the experiments, we first trained MLP and LSTM models without compression as the baseline models.
In the MLP model, three prior frames and the current frame were connected as the input to incorporate the temporal context information and keep it as causal model while only the mask of the current frame was estimated. From the input layer to the output layer, the MLP model had 1024, 1024, 1024, 512, 512, 512, 512, and 256 units, respectively, resulting in 3.54 million parameters in this MLP model. The Rectified Linear Unit (ReLU) activation function was used for the hidden layers while sigmoid activation function was used for the output layer.
The complementary feature without frame expanding and the IRM training target were adopted by the LSTM model. The LSTM model had three LSTM hidden layers with 512 units and one fully-connected layer stacked after the LSTM layers with 256 sigmoid units. The parameter number of this LSTM model was 5.90 million.

During the model training process, the mean square error (MSE) of the estimated mask and the target mask was used as the loss function. The Adam optimizer~\cite{kingma2014adam} was utilized with the learning rate first set to 0.0005 and then decreasing 5\% every 4000 or 1000 training steps respectively for MLP and LSTM models. The dropout technique~\cite{srivastava2014dropout} was employed to avoid potential model overfitting with the dropout rate of 0.3. The MLP and LSTM models were trained with a minibatch size of 1280 at the frame level and 60 at utterance level, respectively. The total number of training epoch was 50.

\subsection{Comparison Methods and Performance Measures}

We compare the proposed MPO-based method with the widely-used pruning method~\cite{han2015deep} at the different compression rates varying from 5 to 100 times. In the pruning method, the unnecessary elements in the weight matrix are eliminated, which means setting these elements to zero to remove the connections with the low weights between the layers of a neural network. During training, the connections with the low weights are removed iteratively based on their magnitude.
In the experiments, both MLP and LSTM models were trained for speech enhancement with the pruning and MPO-based compression methods.

To evaluate the performance of the MPO-based and pruning-based compression methods, they are applied into the MLP and LSTM-based speech enhancement approaches. For the speech enhancement performance evaluation, several objective measures were used, including the short-time objective intelligibility (STOI)~\cite{taal2011algorithm}, the perceptual evaluation of speech quality (PESQ)~\cite{Rix2002Perceptual}, and global SNR.

\section{Experimental Results and Analysis}
\label{sec:result}

We evaluate the MLP and LSTM models before and after compression in two different monaural speech enhancement tasks, \emph{Task \uppercase\expandafter{\romannumeral1}} and \emph{Task \uppercase\expandafter{\romannumeral2}}, as described in Section \ref{subsec:data} . In both tasks, the speech enhancement performances with the pruning- and MPO-based compression models are compared at different compression rates.

\subsection{Evaluation of the Compressed MLP Models}


\subsubsection{Detailed Settings of the MPO and Pruning-based Compression}
\label{setting MLP}
In the pruning method, the sparsity of weight matrices is set to get the compression rate of an MLP model. However, in MPO-based method, there is no such simple setting to get the given compression rate, as parameters $I_k, J_k$ and $D_k$ described in Section \ref{sec:mpomlp} are all adjustable. For convenience, we only tune $D_k$ to get the given compression rate with the fixed $I_k$ and $J_k$. There are four different weight matrix shapes in the MLP model structure. We used the same MPO decomposition format for the weight matrices with the same shape. The fixed $I_k$ and $J_k$ for the different weight matrices are shown in Table \ref{tab:IJmlp}. The values of the tunable bond dimension $D_k$ at different model compression rate are shown in Table \ref{tab:Dmlp}.

\begin{table}
\centering \renewcommand{\arraystretch}{1.1}
\caption{The fixed $I_k$ and $J_k$ in MPO decomposition for the four different weight matrix shapes in the MLP model structure}
\label{tab:IJmlp}
    \begin{tabular}{|c|c|c|}\hline
    \begin{tabular}[c]{@{}c@{}}Weight Matrix\\ $N_h \times N_x$ \end{tabular} &
    \begin{tabular}[c]{@{}c@{}}MPO Decomposed Tensor\\ $(I_1,I_2,I_3,I_4) \times (J_1,J_2,J_3,J_4)$ \end{tabular} &
     \begin{tabular}[c]{@{}c@{}}Tunable Bond\\ Dimension D \end{tabular} \\ \hline
    1024$\times$1024 & (4,8,8,4)$\times$(4,8,8,4) & (1,$D^1$,$D^1$,$D^1$,1) \\ \hline
    512$\times$1024 & (4,4,8,4)$\times$(4,8,8,4) & (1,$D^2$,$D^2$,$D^2$,1) \\ \hline
    512$\times$512 & (4,4,8,4)$\times$(4,4,8,4) & (1,$D^3$,$D^3$,$D^3$,1) \\ \hline
    256$\times$512 & (4,4,4,4)$\times$(4,4,8,4) & (1,$D^4$,$D^4$,$D^4$,1) \\ \hline
    \end{tabular}
\end{table}

\begin{table}
\centering \renewcommand{\arraystretch}{1.1}
\caption{The value of the tunable bond dimension D in different model compression rate in the MLP models}
\label{tab:Dmlp}
    \begin{tabular}{|c|c|c|c|c|c|c|c|c|}\hline
        Compression Rate & 5 & 10 & 15 & 20 & 25 & 50 & 75 & 100 \\ \hline
        $D^1$ & 32 & 23 & 19 & 16 & 15 & 10 & 8 & 7 \\
        $D^2$ & 32 & 23 & 19 & 18 & 13 & 12 & 10 & 8 \\
        $D^3$ & 34 & 23 & 19 & 16 & 15 & 10 & 8 & 7 \\
        $D^4$ & 36 & 23 & 19 & 18 & 15 & 10 & 9 & 8 \\ \hline
    \end{tabular}
\end{table}

\subsubsection{Evaluation Results in Speech Enhancement}

\begin{table*}[!htb]
\centering \renewcommand{\arraystretch}{1.1}
\caption{The evaluation results of the speech enhancement performance for the MLP models with IRM estimation averaged among noise types in three SNR (-5, 0 and 5) environments}
\label{tab:dnn_irm}
    \begin{tabular}{|c|c|ccc|c|ccc|c|ccc|c|}\hline
            \multirow{2}{*}{\begin{tabular}[c]{@{}c@{}}Compression\\ Rate\end{tabular}} & \multirow{2}{*}{\begin{tabular}[c]{@{}c@{}}Compression\\ Method\end{tabular}} & \multicolumn{4}{c|}{STOI} & \multicolumn{4}{c|}{PESQ} & \multicolumn{4}{c|}{SNR} \\ \cline{3-14}
             &  & -5dB & 0dB & 5dB & Avg & -5dB & 0dB & 5dB & Avg & -5dB & 0dB & 5dB & Avg \\ \hline
            0 & --- & 75.41 & 85.69 & 91.79 & 84.29 & 1.58 & 1.99 & 2.47 & 2.02 & 5.10 & 8.88 & 12.68 & 8.89 \\ \hline
            \multirow{2}{*}{5} & pruning & 74.72 & 85.08 & 91.40 & 83.73 & 1.55 & 1.95 & 2.42 & 1.97 & 4.90 & 8.66 & 12.48 & 8.68 \\
             & mpo & \textbf{75.21} & \textbf{85.56} & \textbf{91.71} & \textbf{84.16} & \textbf{1.59} & \textbf{2.00} & \textbf{2.47} & \textbf{2.02} & \textbf{5.15} & \textbf{8.91} & \textbf{12.73} & \textbf{8.93} \\ \hline
            \multirow{2}{*}{10} & pruning & 74.38 & 84.82 & 91.25 & 83.49 & 1.53 & 1.92 & 2.40 & 1.95 & 4.83 & 8.54 & 12.31 & 8.56 \\
             & mpo & \textbf{75.32} & \textbf{85.57} & \textbf{91.68} & \textbf{84.19} & \textbf{1.59} & \textbf{2.00} & \textbf{2.47} & \textbf{2.02} & \textbf{5.15} & \textbf{8.93} & \textbf{12.75} & \textbf{8.94} \\ \hline
            \multirow{2}{*}{15} & pruning & 74.08 & 84.54 & 91.10 & 83.24 & 1.51 & 1.90 & 2.38 & 1.93 & 4.73 & 8.40 & 12.15 & 8.43 \\
             & mpo & \textbf{74.92} & \textbf{85.29} & \textbf{91.56} & \textbf{83.92} & \textbf{1.58} & \textbf{1.99} & \textbf{2.47} & \textbf{2.01} & \textbf{5.11} & \textbf{8.88} & \textbf{12.72} & \textbf{8.91} \\ \hline
            \multirow{2}{*}{20} & pruning & 73.86 & 84.31 & 90.94 & 83.04 & 1.50 & 1.88 & 2.36 & 1.92 & 4.68 & 8.33 & 12.04 & 8.35 \\
             & mpo & \textbf{75.03} & \textbf{85.33} & \textbf{91.51} & \textbf{83.96} & \textbf{1.58} & \textbf{1.98} & \textbf{2.44} & \textbf{2.00} & \textbf{5.09} & \textbf{8.85} & \textbf{12.67} & \textbf{8.87} \\ \hline
            \multirow{2}{*}{25} & pruning & 73.72 & 84.25 & 90.89 & 82.95 & 1.50 & 1.88 & 2.36 & 1.91 & 4.68 & 8.33 & 12.05 & 8.35 \\
             & mpo & \textbf{74.83} & \textbf{85.17} & \textbf{91.45} & \textbf{83.82} & \textbf{1.57} & \textbf{1.97} & \textbf{2.44} & \textbf{1.99} & \textbf{5.05} & \textbf{8.83} & \textbf{12.66} & \textbf{8.85} \\ \hline
            \multirow{2}{*}{50} & pruning & 73.05 & 83.64 & 90.47 & 82.39 & 1.47 & 1.84 & 2.31 & 1.87 & 4.51 & 8.11 & 11.72 & 8.11 \\
             & mpo & \textbf{74.27} & \textbf{84.73} & \textbf{91.17} & \textbf{83.39} & \textbf{1.54} & \textbf{1.94} & \textbf{2.40} & \textbf{1.96} & \textbf{4.97} & \textbf{8.73} & \textbf{12.56} & \textbf{8.75} \\ \hline
            \multirow{2}{*}{75} & pruning & 72.31 & 83.11 & 90.15 & 81.86 & 1.45 & 1.80 & 2.26 & 1.84 & 4.41 & 7.92 & 11.39 & 7.90 \\
             & mpo & \textbf{74.51} & \textbf{84.90} & \textbf{91.27} & \textbf{83.56} & \textbf{1.56} & \textbf{1.95} & \textbf{2.42} & \textbf{1.98} & \textbf{5.07} & \textbf{8.82} & \textbf{12.62} & \textbf{8.83} \\ \hline
            \multirow{2}{*}{100} & pruning & 71.65 & 82.49 & 89.73 & 81.29 & 1.43 & 1.77 & 2.23 & 1.81 & 4.28 & 7.71 & 11.03 & 7.68 \\
             & mpo & \textbf{74.31} & \textbf{84.79} & \textbf{91.21} & \textbf{83.44} & \textbf{1.55} & \textbf{1.95} & \textbf{2.42} & \textbf{1.97} & \textbf{4.91} & \textbf{8.75} & \textbf{12.61} & \textbf{8.76} \\ \hline
            \multicolumn{2}{|c|}{Noisy Speech} & 65.25 & 75.78 & 84.57 & 75.20 & 1.25 & 1.41 & 1.67 & 1.44 & -5.00 & 0.00 & 5.00 &0.00 \\ \hline
    \end{tabular}
\end{table*}

\begin{figure*}[!htb]
    \centering
    \includegraphics[width=\linewidth]{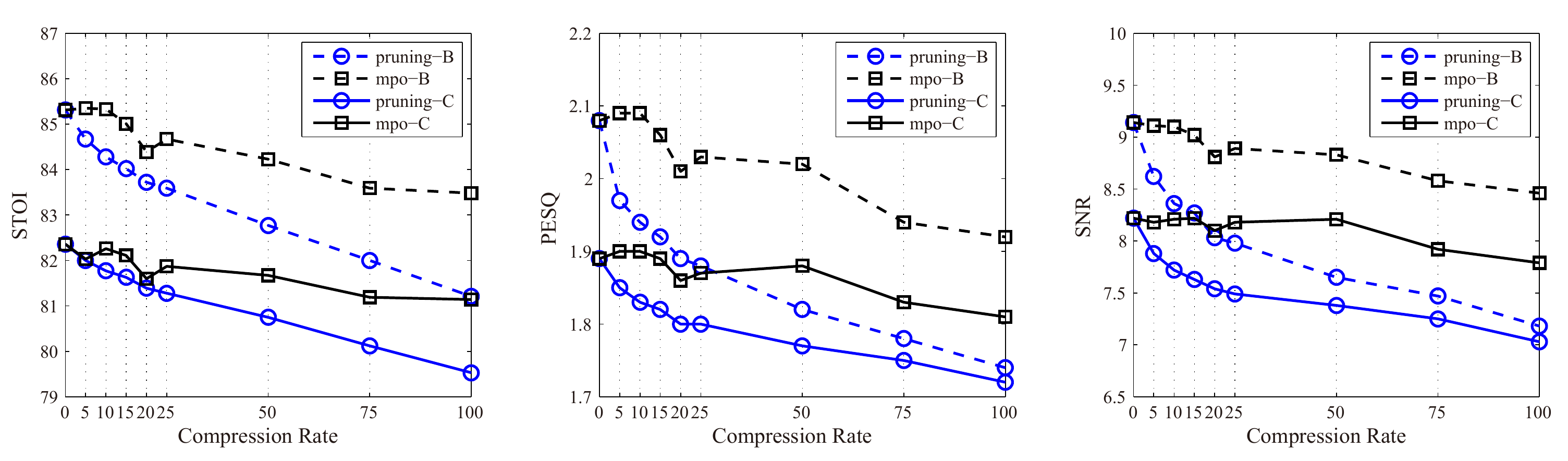}
    \caption{The evaluation results of the speech enhancement performance for the MLP models with IRM estimation in noise dataset B and C averaged among noise types and SNRs.}
    \label{fig:mlp_irm}
\end{figure*}

For the MLP models, the evaluation results of the speech enhancement in \emph{Task \uppercase\expandafter{\romannumeral1}} are shown in Table \ref{tab:dnn_irm}. As we see, the MLP model without compression gains 9.09\% STOI, 0.58 PESQ, and 8.89 dB SNR improvements on average at all the mixing SNR levels in comparison with the noisy speech. These results confirm the effectiveness of the MLP model in monaural speech enhancement task. In terms of the compressed MLP models, the MPO-based compression method outperforms the pruning method at all compression rates. With the pruning methods, the speech enhancement performance decreases with the increasing of compression rate. In contrast, the MPO-based method shows the relatively small performance decrease. However, when the compression rate is small (e.g., 5, 10, 15), the performance of the MLP models with MPO-based compression gains further improvement in PESQ and SNR. These results demonstrate that the MPO-based compression method can be considered as a regularization for the model to reduce the redundance parameters. In the cases with the high compression rates, the MPO-based method can still achieve satisfactory performance. For example,  only 0.85\% STOI, 0.05 PESQ, and 0.14 dB SNR performance losses are respectively introduced when the compression rate is 100 where 35.4 thousands parameters are involved in this compressed model.

Figure \ref{fig:mlp_irm} shows the evaluation results of the speech enhancement for the MLP models in \emph{Task \uppercase\expandafter{\romannumeral2}}. The STOI, PESQ, and SNR results are shown in the left, middle and right panels, respectively. We evaluate the compressed models by both compression methods respectively in the matched noise environments using noise dataset B, referred as \emph{pruning-B} and \emph{mpo-B}, and in the mismatched noise environments using noise dataset C, referred as \emph{pruning-C} and \emph{mpo-C}. The results are averaged over all the noises in each noise dataset and all mixing SNR levels. We see from the results that the MLP models without compression gains satisfactory speech enhancement performance in the mismatched noise condition. The performance of the compressed models in the matched noise condition is better than the one in the mismatched condition for both compression methods at the same compression rate. With the increase of compression rate, the performance of both methods decreases. However, the decrease for the MPO-based method is much slower than for the pruning method. The performance of MPO-based method outperforms the pruning method at the same compression rate.


\subsection{Evaluation of the Compressed LSTM Models}


\subsubsection{Detailed Settings of the MPO and Pruning-based Compression}
\label{setting LSTM}
\begin{table*}[!htb]
\centering \renewcommand{\arraystretch}{1.1}
\caption{The different MPO decomposition solutions for different layers in LSTM model structure} 
\label{tab:IJlstm}
    \begin{tabular}{|c|c|c|c|c|c|c|} \hline
        Solution &Layer &Weight Matrix W& MPO Decomposed Tensor&Weight Matrix U& MPO Decomposed Tensor& Bond Dimension \\ \hline
        \multirow{4}{*}{A} & LSTM1 & 2048$\times$256 & (16,128)$\times$(4,64) & 2048$\times$512 & (16,128)$\times$(4,128) & (1,$D^1$,1) \\
         & LSTM2 & 2048$\times$512 & (16,128)$\times$(4,128) & 2048$\times$512 & (16,128)$\times$(4,128) & (1,$D^2$,1) \\
         & LSTM3 & 2048$\times$512 & (16,128)$\times$(4,128) & 2048$\times$512 & (16,128)$\times$(4,128) & (1,$D^3$,1) \\ \cline{2-7}
         & FC & 256$\times$512 & (4,64)$\times$(4,128) & --- & --- & (1,$D^4$,1) \\ \hline\hline

        \multirow{4}{*}{B} & LSTM1 & 2048$\times$256 & (64,32)$\times$(16,16) & 2048$\times$512 & (64,32)$\times$(16,32) & (1,$D^1$,1) \\
         & LSTM2 & 2048$\times$512 & (64,32)$\times$(16,32) & 2048$\times$512 & (64,32)$\times$(16,32) & (1,$D^2$,1) \\
         & LSTM3 & 2048$\times$512 & (64,32)$\times$(16,32) & 2048$\times$512 & (64,32)$\times$(16,32) & (1,$D^3$,1) \\ \cline{2-7}
         & FC & 256$\times$512 & (16,16)$\times$(16,32) & --- & --- & (1,$D^4$,1) \\ \hline\hline

         \multirow{4}{*}{C}&  LSTM1 & 2048$\times$256 & (8,8,8,4)$\times$(4,4,4,4) & 2048$\times$512 & (8,8,8,4)$\times$(4,4,8,4) & (1,$D^1$,$D^1$,$D^1$,1) \\
        &LSTM2 & 2048$\times$512 & (8,8,8,4)$\times$(4,4,8,4) & 2048$\times$512 & (8,8,8,4)$\times$(4,4,8,4) & (1,$D^2$,$D^2$,$D^2$,1) \\
        &LSTM3 & 2048$\times$512 & (8,8,8,4)$\times$(4,4,8,4) & 2048$\times$512 & (8,8,8,4)$\times$(4,4,8,4) & (1,$D^3$,$D^3$,$D^3$,1) \\ \cline{2-7}
        &FC & 256$\times$512 & (4,4,4,4)$\times$(4,4,8,4) & --- & --- & (1,$D^4$,$D^4$,$D^4$,1) \\ \hline
    \end{tabular}
\end{table*}

As described in Section \ref{subsec:mpllstm}, there are two weight matrices in an LSTM layer, the input-to-hidden weight matrix \textbf{W} and the hidden-to-hidden weight matrix \textbf{U}. In MPO-based model compression method, both weight matrices are decomposed. Therefore, the parameters including factorization factors $I_k^W$, $J_k^W, I_k^U$, $J_k^U$ and bond dimension factors $D_k^W, D_k^U$ are adjustable. Similar to the case with the MLP models, we fixed the factorization factors while three decomposition solutions were used at different compression rate. The decomposition solutions are shown in Table \ref{tab:IJlstm}, in which layers LSTM1, LSTM2, LSTM3, and FC are the first, second, third LSTM layers, and the one fully-connected layer, respectively. The bond dimension factors for the two weight matrices in an LSTM layer are identical to each other and tuned to achieve a given compression rate. The decomposition solution and the value of the tunable bond dimension at different model compression rates can be found in Table \ref{tab:Dlstm}.
In this paper, we have done the decomposition just by convenience. In fact, it can be argued and veriﬁed by examples that when the network is away from under-ﬁtting, diﬀerent factorization manners should always produce almost the same result as discussed in the supplemental material of \cite{gao2019compressing}.
In the pruning method, we only need to determine the sparsity of weight matrix which is the same as the case with the MLP model.
\begin{table}[!htb]
\centering \renewcommand{\arraystretch}{1.1}
\caption{The decomposition solutions and the tunable bond dimension values at different model compression rates for the LSTM models}
\label{tab:Dlstm}
    \begin{tabular}{|c|c|c|c|c|c|c|c|c|}\hline
        Compression Rate & 5 & 10 & 15 & 20 & 25 & 50 & 75 & 100 \\ \hline
        Solution & A & B & B & B & C & C & C & C \\ \hline
        $D^1$ & 14 & 47 & 31 & 24 & 20 & 14 & 12 & 10 \\
        $D^2$ & 14 & 47 & 31 & 24 & 20 & 14 & 11 & 9 \\
        $D^3$ & 14 & 47 & 31 & 24 & 20 & 14 & 11 & 10 \\
        $D^4$ & 12 & 47 & 31 & 24 & 20 & 13 & 11 & 9 \\ \hline
    \end{tabular}
\end{table}

\subsubsection{Evaluation Results in Speech Enhancement}

For the LSTM models, the evaluation results of the speech enhancement in \emph{Task \uppercase\expandafter{\romannumeral1}} are shown in Table \ref{tab:lstmirm}. These results also confirm the effectiveness of the LSTM model in monaural speech enhancement task. Actually, in comparison, it outperforms the MLP model. We see from the results that the LSTM model without compression respectively gains 10.8\% STOI, 0.77 PESQ, and 9.34 dB SNR improvements on average of all mixing SNR levels in comparison with the noisy speech. In terms of compressed models, the speech enhancement performance decreases with the increase of compression rate for both compression methods. Again, the MPO-based compression method outperforms the pruning method in all evaluation measures and compression rates. In the case with the high compression rate, for example 100 times, the compressed model with the MPO-based compression method can still achieve satisfactory performance respectively with only 1.97\% STOI, 0.2 PESQ, and 0.5 dB SNR losses with only 59.0 thousands parameters.

\begin{figure*}[!htb]
    \centering
    \includegraphics[width=\linewidth]{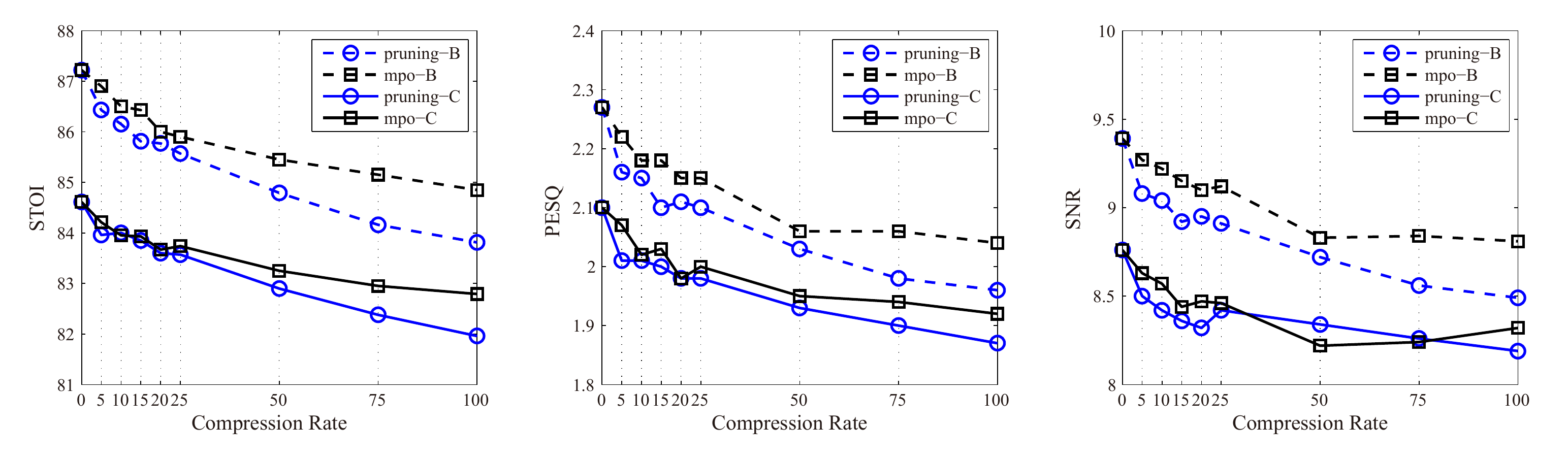}
    \caption{The evaluation results of the speech enhancement performance for the LSTM models with IRM estimation in noise dataset B and C averaged among noise types and SNRs.}
    \label{fig:lstm_irm}
\end{figure*}

The evaluation results of the speech enhancement for the LSTM models in \emph{Task \uppercase\expandafter{\romannumeral2}} is shown in Fig.\ref{fig:lstm_irm}. The STOI, PESQ, and SNR results are shown in the left, middle and right panels, respectively. The results of both compression methods in the matched and mismatched noise environments are plotted in \emph{pruning-B}, \emph{mpo-B}, \emph{pruning-C}, and \emph{mpo-C}. The results are averaged among the noises in each noise dataset and all mixing SNR levels. Similar to the case of \emph{Task \uppercase\expandafter{\romannumeral1}}, the LSTM model without compression obtains satisfactory speech enhancement performance in the matched and mismatched noise environments and outperforms the MLP model. However, there is still a gap between the performance in mismatched and matched noise conditions. In terms of compressed models, the advantage of the MPO-based compression method is readily shown in comparison with the pruning method at the same compression rate. As the number of parameters in the LSTM model is lager than that in the MLP model and the LSTM model gains better speech enhancement performance, the advantage of the MPO-based compression method in comparison with the pruning method for the LSTM model is not so significant as for the MLP model. In some cases, such as SNR evaluation at a compression rate of 50 times in the mismatched noise condition, the pruning method shows better performance. Nevertheless, the MPO-based method outperforms the pruning method overall. The effectiveness of the MPO-based compression method for LSTM model in speech enhancement can be confirmed.

\begin{table*}[!htb]
\centering \renewcommand{\arraystretch}{1.1}
\caption{The evaluation results of the speech enhancement performance for the LSTM models with IRM estimation averaged among noise types in three SNR (-5, 0 and 5) environments}
\label{tab:lstmirm}
    \begin{tabular}{|c|c|c|c|c|c|c|c|c|c|c|c|c|c|}\hline
        \multirow{2}{*}{\begin{tabular}[c]{@{}c@{}}Compression\\ Rate\end{tabular}} & \multirow{2}{*}{\begin{tabular}[c]{@{}c@{}}Compression\\ Method\end{tabular}} & \multicolumn{4}{c|}{STOI} & \multicolumn{4}{c|}{PESQ} & \multicolumn{4}{c|}{SNR} \\ \cline{3-14}
         &  & -5dB & 0dB & 5dB & Avg & -5dB & 0dB & 5dB & Avg & -5dB & 0dB & 5dB & Avg \\ \hline
        0 & - & 78.00 & 87.38 & 92.62 & 86.00 & 1.74 & 2.21 & 2.68 & 2.21 & 5.66 & 9.32 & 13.02 & 9.34 \\ \hline
        \multirow{2}{*}{5} & pruning & 76.73 & 86.35 & 91.99 & 85.02 & 1.66 & 2.08 & 2.54 & 2.09 & 5.37 & 8.98 & 12.59 & 8.98 \\
         & mpo & \textbf{77.75} & \textbf{87.23} & \textbf{92.50} & \textbf{85.83} & \textbf{1.72} & \textbf{2.16} & \textbf{2.63} & \textbf{2.17} & \textbf{5.60} & \textbf{9.25} & \textbf{12.96} & \textbf{9.27} \\ \hline
        \multirow{2}{*}{10} & pruning & 76.43 & 86.11 & 91.82 & 84.79 & 1.64 & 2.06 & 2.51 & 2.07 & 5.36 & 8.94 & 12.56 & 8.95 \\
         & mpo & \textbf{76.82} & \textbf{86.56} & \textbf{92.15} & \textbf{85.17} & \textbf{1.68} & \textbf{2.11} & \textbf{2.57} & \textbf{2.12} & \textbf{5.54} & \textbf{9.16} & \textbf{12.87} & \textbf{9.19} \\ \hline
        \multirow{2}{*}{15} & pruning & 76.31 & 86.00 & 91.76 & 84.69 & 1.63 & 2.04 & 2.50 & 2.06 & 5.32 & 8.91 & 12.52 & 8.92 \\
         & mpo & \textbf{76.65} & \textbf{86.45} & \textbf{92.06} & \textbf{85.06} & \textbf{1.68} & \textbf{2.10} & \textbf{2.56} & \textbf{2.11} & \textbf{5.50} & \textbf{9.12} & \textbf{12.85} & \textbf{9.16} \\ \hline
        \multirow{2}{*}{20} & pruning & 75.99 & 85.77 & 91.62 & 84.46 & 1.62 & 2.02 & 2.48 & 2.04 & 5.29 & 8.86 & 12.48 & 8.88 \\
         & mpo & \textbf{76.52} & \textbf{86.34} & \textbf{92.02} & \textbf{84.96} & \textbf{1.66} & \textbf{2.07} & \textbf{2.53} & \textbf{2.09} & \textbf{5.40} & \textbf{9.08} & \textbf{12.83} & \textbf{9.10} \\ \hline
        \multirow{2}{*}{25} & pruning & 75.80 & 85.60 & 91.53 & 84.31 & 1.61 & 2.00 & 2.46 & 2.02 & 5.24 & 8.82 & 12.44 & 8.84 \\
         & mpo & \textbf{76.56} & \textbf{86.27} & \textbf{91.94} & \textbf{84.92} & \textbf{1.65} & \textbf{2.06} & \textbf{2.51} & \textbf{2.08} & \textbf{5.35} & \textbf{8.97} & \textbf{12.70} & \textbf{9.01} \\ \hline
        \multirow{2}{*}{50} & pruning & 74.87 & 84.94 & 91.15 & 83.66 & 1.57 & 1.95 & 2.40 & 1.98 & 5.09 & 8.67 & 12.30 & 8.68 \\
         & mpo & \textbf{76.03} & \textbf{85.77} & \textbf{91.65} & \textbf{84.49} & \textbf{1.62} & \textbf{2.02} & \textbf{2.46} & \textbf{2.03} & \textbf{5.21} & \textbf{8.85} & \textbf{12.62} & \textbf{8.89} \\ \hline
        \multirow{2}{*}{75} & pruning & 74.24 & 84.46 & 90.87 & 83.19 & 1.54 & 1.91 & 2.35 & 1.93 & 4.94 & 8.54 & 12.15 & 8.54 \\
         & mpo & \textbf{75.41} & \textbf{85.39} & \textbf{91.42} & \textbf{84.07} & \textbf{1.59} & \textbf{1.98} & \textbf{2.43} & \textbf{2.00} & \textbf{5.14} & \textbf{8.78} & \textbf{12.56} & \textbf{8.83} \\ \hline
        \multirow{2}{*}{100} & pruning & 73.65 & 84.02 & 90.63 & 82.77 & 1.51 & 1.87 & 2.32 & 1.90 & 4.84 & 8.41 & 12.04 & 8.43 \\
         & mpo & \textbf{75.32} & \textbf{85.36} & \textbf{91.42} & \textbf{84.03} & \textbf{1.59} & \textbf{1.99} & \textbf{2.45} & \textbf{2.01} & \textbf{5.14} & \textbf{8.80} & \textbf{12.57} & \textbf{8.84} \\ \hline
        \multicolumn{2}{|c|}{Noisy Speech} & 65.25 & 75.78 & 84.57 & 75.20 & 1.25 & 1.41 & 1.67 & 1.44 & -5.00 & 0.00 & 5.00 &0.00 \\ \hline
    \end{tabular}
\end{table*}

\subsection{Computational Complexity Analysis of the Compressed Model}

\begin{figure}[!htb]
    \centering
    \includegraphics[width=0.8\linewidth]{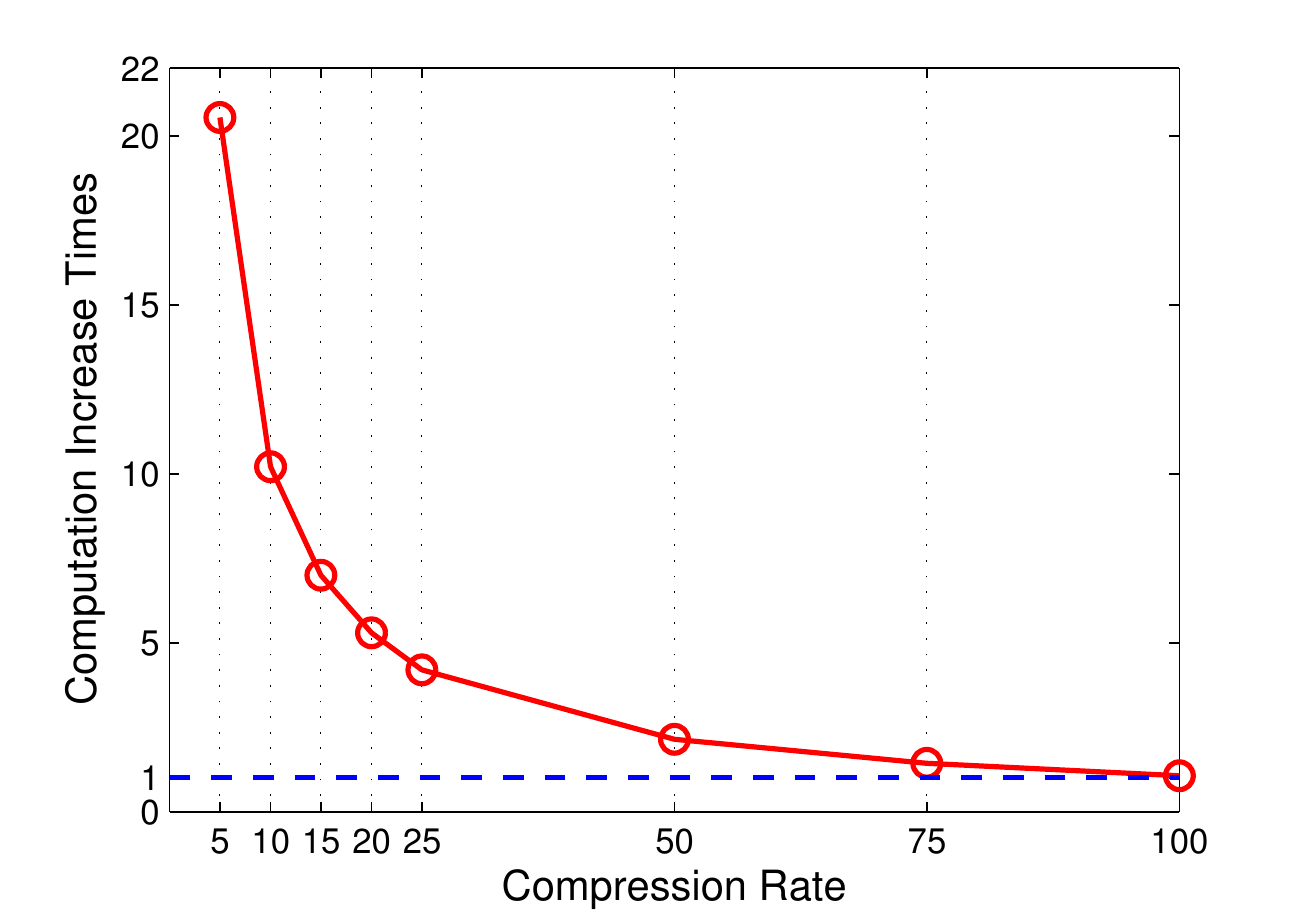}
    \caption{The computation increase times of the compressed MLP models in different compression rates.}
    \label{fig:dnn_comp}
\end{figure}

\begin{figure}[!htb]
    \centering
    \includegraphics[width=0.8\linewidth]{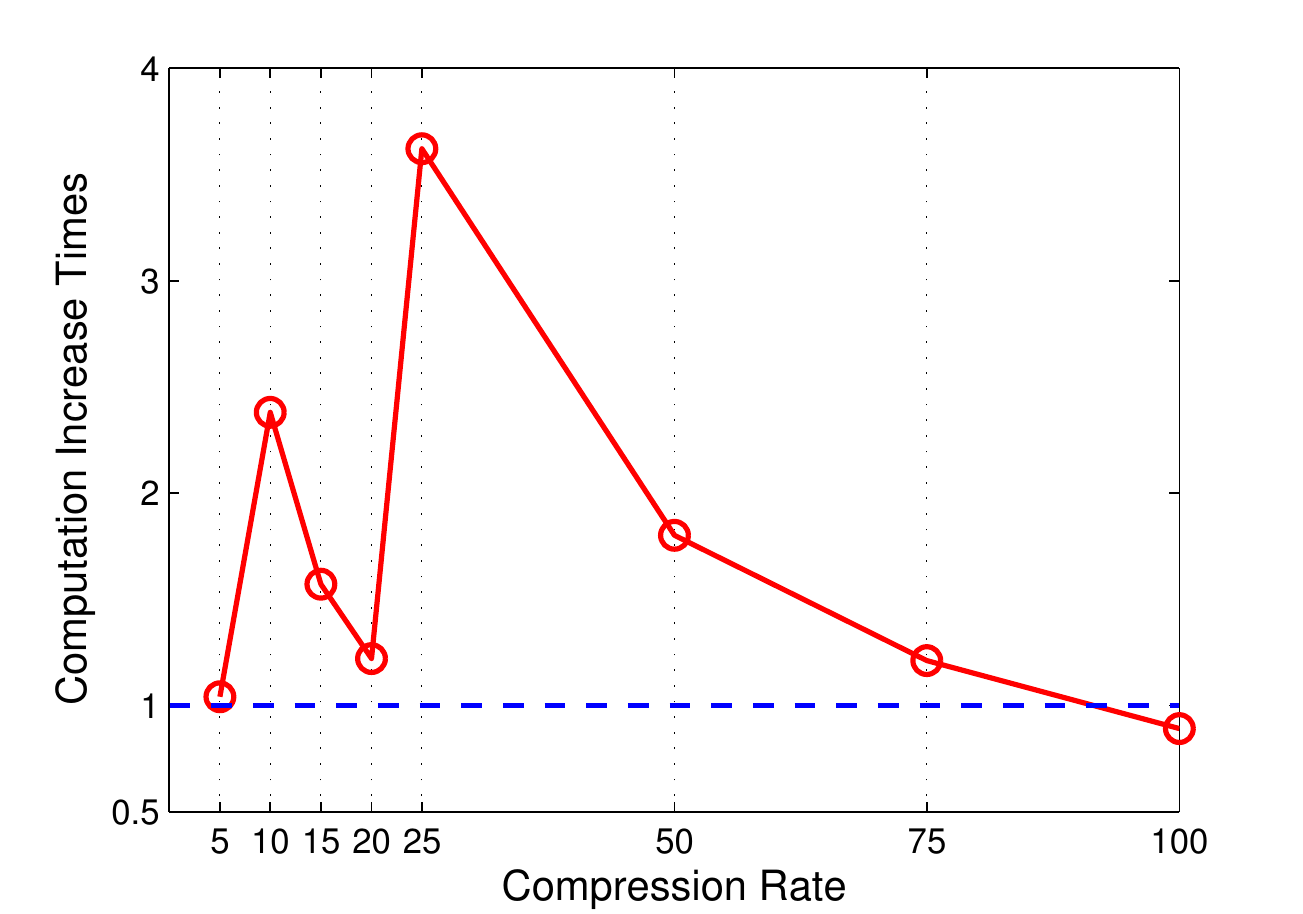}
    \caption{The computation increase times of the compressed LSTM models in different compression rates.}
    \label{fig:lstm_comp}
\end{figure}

Except the reduction of model parameters, the computational complexity is also an important aspect in the real application of speech enhancement. In this subsection, we calculate the number of multiplication operators in the testing stage of the uncompressed and compressed models.
We plot the compression rate against the increase times of computational cost for both compressed MLP and LSTM models, as shown in Figs. \ref{fig:dnn_comp} and \ref{fig:lstm_comp}.
As described in Section~\ref{setting MLP}, the number of local tensor in MPO decomposition format of the MLP model is fixed and the compression rate is increased with the decrease of the bond dimension.
As shown in Fig.\ref{fig:dnn_comp}, the computation increase times of the compressed MLP model decrease with the increase of compression rates. In the compression rate of about 100 times, the computational complexity approaches that of the uncompressed model.
Different from the compression of MLP model, we use three different MPO decomposition solutions for LSTM model compression as described in Section~\ref{setting LSTM}.
The computation increase times of the compressed LSTM model vary in different solutions. Meanwhile, in the same solution, the computation increase times decreases with the increase of compression rates, such as from the compression rate of 25 to 100 times in which the solution C is used. In LSTM model compression, the computational complexity approaches that of the uncompressed model with the compression rates of 5, 20, 75 and 100 times and there is no large increase overall. These analysis means that the MPO decomposition format can be carefully selected to maintain or reduce the computational complexity as well as compress the quantity of model parameters.

\section{Conclusion and Discussion}
\label{sec:conclusion}

In this paper, we propose a novel model compression method based on matrix product operators to reduce model size and apply it into the DNN-based monaural speech enhancement task. In the MPO-based compression method, we replace the weight matrices using the MPO decomposition format in the linear transformations of the MLP and LSTM models. We evaluate the models at different compression rates for the IRM-based speech enhancement. The experimental results show that our proposed MPO-based method outperforms the pruning method in terms of speech enhancement performance at the same compression rate. Meanwhile, further improvement can be obtained by the MPO-based compression at a low compression rate, which means it can be considered as regularization process to reduce the redundant parameters.
In comparison with the other methods, the MPO-based model compression method is more suitable for a platform possessing the limited memory resource and the high computational ability for tensor product, especially in cloud-free applications, as there is a sequence of multiplications for high-dimensional tensor in the compressed model. However, the computational complexity varies in different MPO decomposition solutions and decreased with the compression rate in the same solution. The MPO decomposition format can be carefully selected to maintain or reduce the computation complexity as while as compress the number of model parameters.
Furthermore, the MPO-based model compression method can be readily integrated to any network models with the linear transformations and trained end-to-end to work together with the rest of models in a very efficient way.

\section*{Acknowledgment}
This research is partially supported by the National Natural Science Foundation of China (Nos. 11934020, 11722437, 11674352, 11774422).

\ifCLASSOPTIONcaptionsoff
  \newpage
\fi



%
%
%
\bibliographystyle{IEEEtran}
\bibliography{refs}

\begin{thebibliography}{10}
\providecommand{\url}[1]{#1}
\csname url@samestyle\endcsname
\providecommand{\newblock}{\relax}
\providecommand{\bibinfo}[2]{#2}
\providecommand{\BIBentrySTDinterwordspacing}{\spaceskip=0pt\relax}
\providecommand{\BIBentryALTinterwordstretchfactor}{4}
\providecommand{\BIBentryALTinterwordspacing}{\spaceskip=\fontdimen2\font plus
\BIBentryALTinterwordstretchfactor\fontdimen3\font minus
  \fontdimen4\font\relax}
\providecommand{\BIBforeignlanguage}[2]{{%
\expandafter\ifx\csname l@#1\endcsname\relax
\typeout{** WARNING: IEEEtran.bst: No hyphenation pattern has been}%
\typeout{** loaded for the language `#1'. Using the pattern for}%
\typeout{** the default language instead.}%
\else
\language=\csname l@#1\endcsname
\fi
#2}}
\providecommand{\BIBdecl}{\relax}
\BIBdecl

\bibitem{loizou2007speech}
P.~C. Loizou, \emph{Speech enhancement: theory and practice}.\hskip 1em plus
  0.5em minus 0.4em\relax CRC press, 2007.

\bibitem{boll1979suppression}
S.~Boll, ``Suppression of acoustic noise in speech using spectral
  subtraction,'' \emph{IEEE Transactions on acoustics, speech, and signal
  processing}, vol.~27, no.~2, pp. 113--120, 1979.

\bibitem{Lim1979Enhancement}
J.~S. {Lim} and A.~V. {Oppenheim}, ``Enhancement and bandwidth compression of
  noisy speech,'' \emph{Proceedings of the IEEE}, vol.~67, no.~12, pp.
  1586--1604, Dec 1979.

\bibitem{ephraim1984speech}
Y.~Ephraim and D.~Malah, ``Speech enhancement using a minimum-mean square error
  short-time spectral amplitude estimator,'' \emph{IEEE Transactions on
  acoustics, speech, and signal processing}, vol.~32, no.~6, pp. 1109--1121,
  1984.

\bibitem{wang2018supervised}
D.~Wang and J.~Chen, ``Supervised speech separation based on deep learning: An
  overview,'' \emph{IEEE/ACM Transactions on Audio, Speech, and Language
  Processing}, 2018.

\bibitem{xu2014experimental}
Y.~Xu, J.~Du, L.-R. Dai, and C.-H. Lee, ``An experimental study on speech
  enhancement based on deep neural networks,'' \emph{IEEE Signal processing
  letters}, vol.~21, no.~1, pp. 65--68, 2014.

\bibitem{xu2015regression}
------, ``A regression approach to speech enhancement based on deep neural
  networks,'' \emph{IEEE/ACM Transactions on Audio, Speech and Language
  Processing (TASLP)}, vol.~23, no.~1, pp. 7--19, 2015.

\bibitem{wang2013towards}
Y.~Wang and D.~Wang, ``Towards scaling up classification-based speech
  separation,'' \emph{IEEE Transactions on Audio, Speech, and Language
  Processing}, vol.~21, no.~7, pp. 1381--1390, 2013.

\bibitem{wang2005ideal}
D.~Wang, ``On ideal binary mask as the computational goal of auditory scene
  analysis,'' in \emph{Speech separation by humans and machines}.\hskip 1em
  plus 0.5em minus 0.4em\relax Springer, 2005, pp. 181--197.

\bibitem{wang2014training}
Y.~Wang, A.~Narayanan, and D.~Wang, ``On training targets for supervised speech
  separation,'' \emph{IEEE/ACM transactions on audio, speech, and language
  processing}, vol.~22, no.~12, pp. 1849--1858, 2014.

\bibitem{erdogan2015phase}
H.~Erdogan, J.~R. Hershey, S.~Watanabe, and J.~Le~Roux, ``Phase-sensitive and
  recognition-boosted speech separation using deep recurrent neural networks,''
  in \emph{2015 IEEE International Conference on Acoustics, Speech and Signal
  Processing (ICASSP)}, 2015, pp. 708--712.

\bibitem{williamson2016complex}
D.~S. Williamson, Y.~Wang, and D.~Wang, ``Complex ratio masking for monaural
  speech separation,'' \emph{IEEE/ACM Transactions on Audio, Speech and
  Language Processing (TASLP)}, vol.~24, no.~3, pp. 483--492, 2016.

\bibitem{kolbk2017speech}
M.~Kolbk, Z.-H. Tan, and J.~Jensen, ``Speech intelligibility potential of
  general and specialized deep neural network based speech enhancement
  systems,'' \emph{IEEE/ACM Transactions on Audio, Speech and Language
  Processing (TASLP)}, vol.~25, no.~1, pp. 153--167, 2017.

\bibitem{chen2017long}
J.~Chen and D.~Wang, ``Long short-term memory for speaker generalization in
  supervised speech separation,'' \emph{The Journal of the Acoustical Society
  of America}, vol. 141, no.~6, pp. 4705--4714, 2017.

\bibitem{weninger2014single}
F.~Weninger, F.~Eyben, and B.~Schuller, ``Single-channel speech separation with
  memory-enhanced recurrent neural networks,'' in \emph{2014 IEEE International
  Conference on Acoustics, Speech and Signal Processing (ICASSP)}.\hskip 1em
  plus 0.5em minus 0.4em\relax IEEE, 2014, pp. 3709--3713.

\bibitem{weninger2015speech}
F.~Weninger, H.~Erdogan, S.~Watanabe, E.~Vincent, J.~Le~Roux, J.~R. Hershey,
  and B.~Schuller, ``Speech enhancement with lstm recurrent neural networks and
  its application to noise-robust asr,'' in \emph{International Conference on
  Latent Variable Analysis and Signal Separation}.\hskip 1em plus 0.5em minus
  0.4em\relax Springer, 2015, pp. 91--99.

\bibitem{park2016fully}
S.~R. Park and J.~Lee, ``A fully convolutional neural network for speech
  enhancement,'' \emph{arXiv preprint arXiv:1609.07132}, 2016.

\bibitem{fu2017raw}
S.-W. Fu, Y.~Tsao, X.~Lu, and H.~Kawai, ``Raw waveform-based speech enhancement
  by fully convolutional networks,'' in \emph{2017 Asia-Pacific Signal and
  Information Processing Association Annual Summit and Conference (APSIPA
  ASC)}.\hskip 1em plus 0.5em minus 0.4em\relax IEEE, 2017, pp. 006--012.

\bibitem{pandey2019tcnn}
A.~Pandey and D.~Wang, ``Tcnn: Temporal convolutional neural network for
  real-time speech enhancement in the time domain,'' in \emph{ICASSP 2019-2019
  IEEE International Conference on Acoustics, Speech and Signal Processing
  (ICASSP)}.\hskip 1em plus 0.5em minus 0.4em\relax IEEE, 2019, pp. 6875--6879.

\bibitem{hinton2015distilling}
G.~Hinton, O.~Vinyals, and J.~Dean, ``Distilling the knowledge in a neural
  network,'' \emph{stat}, vol. 1050, p.~9, 2015.

\bibitem{ba2014deep}
J.~Ba and R.~Caruana, ``Do deep nets really need to be deep?'' in
  \emph{Advances in neural information processing systems}, 2014, pp.
  2654--2662.

\bibitem{tang2016recurrent}
Z.~Tang, D.~Wang, and Z.~Zhang, ``Recurrent neural network training with dark
  knowledge transfer,'' in \emph{2016 IEEE International Conference on
  Acoustics, Speech and Signal Processing (ICASSP)}.\hskip 1em plus 0.5em minus
  0.4em\relax IEEE, 2016, pp. 5900--5904.

\bibitem{han2015deep}
\BIBentryALTinterwordspacing
S.~Han, H.~Mao, and W.~J. Dally, ``Deep compression: Compressing deep neural
  network with pruning, trained quantization and huffman coding,'' in \emph{4th
  International Conference on Learning Representations, {ICLR} 2016, San Juan,
  Puerto Rico, May 2-4, 2016, Conference Track Proceedings}, Y.~Bengio and
  Y.~LeCun, Eds., 2016. [Online]. Available:
  \url{http://arxiv.org/abs/1510.00149}
\BIBentrySTDinterwordspacing

\bibitem{han2015learning}
S.~Han, J.~Pool, J.~Tran, and W.~Dally, ``Learning both weights and connections
  for efficient neural network,'' in \emph{Advances in neural information
  processing systems}, 2015, pp. 1135--1143.

\bibitem{chen2015compressing}
W.~Chen, J.~Wilson, S.~Tyree, K.~Weinberger, and Y.~Chen, ``Compressing neural
  networks with the hashing trick,'' in \emph{International Conference on
  Machine Learning}, 2015, pp. 2285--2294.

\bibitem{xue2013restructuring}
J.~Xue, J.~Li, and Y.~Gong, ``Restructuring of deep neural network acoustic
  models with singular value decomposition.'' in \emph{Interspeech}, 2013, pp.
  2365--2369.

\bibitem{denil2013predicting}
M.~Denil, B.~Shakibi, L.~Dinh, M.~Ranzato, and N.~De~Freitas, ``Predicting
  parameters in deep learning,'' in \emph{Advances in neural information
  processing systems}, 2013, pp. 2148--2156.

\bibitem{oseledets2011tensor}
I.~V. Oseledets, ``Tensor-train decomposition,'' \emph{SIAM Journal on
  Scientific Computing}, vol.~33, no.~5, pp. 2295--2317, 2011.

\bibitem{novikov2015tensorizing}
A.~Novikov, D.~Podoprikhin, A.~Osokin, and D.~P. Vetrov, ``Tensorizing neural
  networks,'' in \emph{Advances in neural information processing systems},
  2015, pp. 442--450.

\bibitem{Khrulkov2015tensorized}
M.~L. Khrulkov~V, Hrinchuk~O, ``Tensorized embedding layers for efficient model
  compression,'' \emph{arXiv preprint arXiv:1901.10787v2}, 2019.

\bibitem{gao2019compressing}
\BIBentryALTinterwordspacing
Z.-F. Gao, S.~Cheng, R.-Q. He, Z.~Y. Xie, H.-H. Zhao, Z.-Y. Lu, and T.~Xiang,
  ``Compressing deep neural networks by matrix product operators,'' \emph{Phys.
  Rev. Research}, vol.~2, p. 023300, Jun 2020. [Online]. Available:
  \url{https://link.aps.org/doi/10.1103/PhysRevResearch.2.023300}
\BIBentrySTDinterwordspacing

\bibitem{garipov2016ultimate}
T.~Garipov, D.~Podoprikhin, A.~Novikov, and D.~Vetrov, ``Ultimate
  tensorization: compressing convolutional and fc layers alike,'' \emph{arXiv
  preprint arXiv:1611.03214}, 2016.

\bibitem{Yinchong2017tensor}
V.~T. Yinchong~Y, Denis~K, ``Tensor-train recurrent neural networks for video
  classification,'' \emph{ICML}, vol. 3891, p.~70, 2017.

\bibitem{wang2013exploring}
Y.~{Wang}, K.~{Han}, and D.~{Wang}, ``Exploring monaural features for
  classification-based speech segregation,'' \emph{IEEE Transactions on Audio,
  Speech, and Language Processing}, vol.~21, no.~2, pp. 270--279, Feb 2013.

\bibitem{verstraete2004matrix}
F.~Verstraete, J.~J. Garcia-Ripoll, and J.~I. Cirac, ``Matrix product density
  operators: simulation of finite-temperature and dissipative systems,''
  \emph{Physical review letters}, vol.~93, no.~20, p. 207204, 2004.

\bibitem{poulin2011quantum}
D.~Poulin, A.~Qarry, R.~Somma, and F.~Verstraete, ``Quantum simulation of
  time-dependent hamiltonians and the convenient illusion of hilbert space,''
  \emph{Physical review letters}, vol. 106, no.~17, p. 170501, 2011.

\bibitem{chollet2015keras}
F.~Chollet \emph{et~al.}, ``Keras: Deep learning library for theano and
  tensorflow,'' \emph{URL: https://keras. io/k}, vol.~7, no.~8, p.~T1, 2015.

\bibitem{Andrew1993Assessment}
A.~Varga and H.~J. Steeneken, ``Assessment for automatic speech recognition:
  Ii. noisex-92: A database and an experiment to study the effect of additive
  noise on speech recognition systems,'' \emph{Speech Communication}, vol.~12,
  no.~3, pp. 247 -- 251, 1993.

\bibitem{thiemann2013DEMAND}
J.~Thiemann, N.~Ito, and E.~Vincent, ``Demand: a collection of multi-channel
  recordings of acoustic noise in diverse environments,'' Jun. 2013, {Supported
  by Inria under the Associate Team Program VERSAMUS}.

\bibitem{Barker2015CHiME}
J.~Barker, R.~Marxer, E.~Vincent, and S.~Watanabe, ``The third 'chime' speech
  separation and recognition challenge: Dataset, task and baselines,'' in
  \emph{2015 IEEE Workshop on Automatic Speech Recognition and Understanding
  (ASRU)}.\hskip 1em plus 0.5em minus 0.4em\relax IEEE, 2015, pp. 504--511.

\bibitem{kingma2014adam}
D.~P. Kingma and J.~Ba, ``Adam: A method for stochastic optimization,''
  \emph{ICLR}, 2015.

\bibitem{srivastava2014dropout}
N.~Srivastava, G.~Hinton, A.~Krizhevsky, I.~Sutskever, and R.~Salakhutdinov,
  ``Dropout: a simple way to prevent neural networks from overfitting,''
  \emph{The Journal of Machine Learning Research}, vol.~15, no.~1, pp.
  1929--1958, 2014.

\bibitem{taal2011algorithm}
C.~H. Taal, R.~C. Hendriks, R.~Heusdens, and J.~Jensen, ``An algorithm for
  intelligibility prediction of time--frequency weighted noisy speech,''
  \emph{IEEE Transactions on Audio, Speech, and Language Processing}, vol.~19,
  no.~7, pp. 2125--2136, 2011.

\bibitem{Rix2002Perceptual}
A.~W. Rix, J.~G. Beerends, M.~P. Hollier, and A.~P. Hekstra, ``Perceptual
  evaluation of speech quality (pesq)-a new method for speech quality
  assessment of telephone networks and codecs,'' in \emph{IEEE International
  Conference on Acoustics}, 2002.

\end{thebibliography}
%

%
%
%




\end{document}